# A novel high-current, high-resolution, low-kinetic-energy electron source for inverse photoemission spectroscopy

Harald Ibach[1], Haruki Sato (佐藤晴輝)[2], Mihiro Kubo (久保美潤)[2], F. Stefan Tautz[3,4,5], Hiroyuki Yoshida (吉田弘幸)[2,6,7], François C. Bocquet[3,4,*]

[1]*Peter Grünberg Institut (PGI-6), Forschungszentrum Jülich, 52425 Jülich, Germany*
[2]*Graduate School of Science and Engineering, Chiba University, 1-33 Yayoi-cho, Inage-ku, Chiba 263-8522, Japan*
[3]*Peter Grünberg Institut (PGI-3), Forschungszentrum Jülich, 52425 Jülich, Germany*
[4]*Jülich Aachen Research Alliance (JARA), Fundamentals of Future Information Technology, 52425 Jülich, Germany*
[5]*Experimentalphysik IV A, RWTH Aachen University, Otto-Blumenthal-Strasse, 52074 Aachen, Germany*
[6]*Graduate School of Engineering, Chiba University, 1-33 Yayoi-cho, Inage-ku, Chiba 263-8522, Japan*
[7]*Molecular Chirality Research Center, Chiba University, 1-33 Yayoi-cho, Inage-ku, Chiba 263-8522, Japan*
*Corresponding author: f.bocquet@fz-juelich.de

A high-current electron source for inverse photoemission spectroscopy (IPES) is described. The source comprises a thermal cathode electron emission system, an electrostatic deflector-monochromator, and a lens system for variable kinetic energy (1.6 - 20 eV) at the target. When scaled to the energy resolution, the electron current is an order of magnitude higher than that of previously described electron sources developed in the context of electron energy loss spectroscopy. Surprisingly, the experimentally measured energy resolution turned out to be significantly better than calculated by standard programs, which include the electron-electron repulsion in the continuum approximation. The achieved currents are also significantly higher than predicted. We attribute this "*inverse Boersch-effect*" to a mechanism of velocity selection in the forward direction by binary electron-electron collisions.

## 1. Introduction

The term inverse photoemission spectroscopy (IPES) denotes an experimental technique in which a solid surface is exposed to an electron beam of preferably low kinetic energies and the emitted light is detected [1-3]. Electrons enter the solid with an energy above the vacuum level and may undergo radiative decay into unoccupied energy levels. The intensity of the emitted light scales with the density of the unoccupied states, which is the basis for the spectroscopy. The resolution of the spectroscopy is determined by the energy width of the electron beam and the band width of the light detector. The efficiency of the inverse photoemission process is rather low [4]. Effective light detection and high intensity of the electron beam are therefore required. Using an improved Geiger-Müller counter combined with a toroidal 90° electrostatic deflector, resolution below 200 meV were reported in 2007 [5]. In 2012, a fundamentally new method of light detection in a narrow energy window has been introduced by H. Yoshida [6-8], employing a combination of interference filters and low dark counts photomultipliers, which resulted in a significant improvement over the original set-up of Dose *et al*. [1]. The interference filters select a band of about 50 meV. Instead of interference filters, Czerny-Turner grating spectrometers may also be used for light detection [9]. Presently, the resolution of the technique is limited by the energy width of the electron beam, which is currently determined by the thermal emission process. For low-temperature BaO cathodes (1100K), Yoshida estimates the energy width to 250meV [7]. In order to achieve higher resolution and higher currents the Boersch effect need be considered which increases the energy width in dense, accelerated electron beams [10-14]. The objective of the endeavor presented here is to design and manufacture an electron source that produces the highest possible electron current of smallest possible energy spread at low, variable impact energy on the target surface.

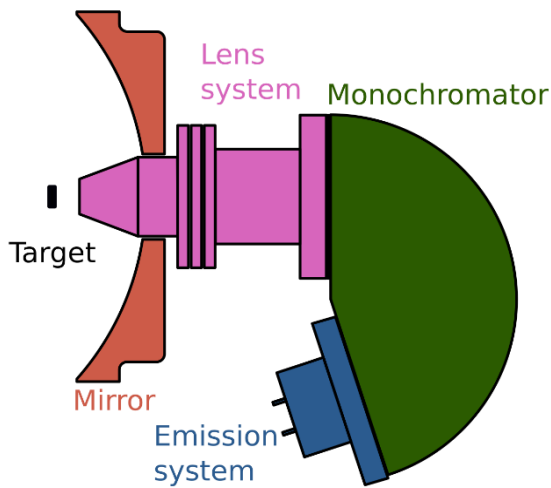

Fig. 1: Overview of the electron source, installed through a typical elliptical mirror used for IPES.

The electron source consists of three units, as seen in Fig. 1. The first one is the thermal cathode electron emission system, which delivers the input current for the monochromator. The second unit is an electrostatic deflector serving as monochromator. The final unit is a lens system that allows for variable, low impact energies at the target. Each of the three units must match the properties of the other two. For example, electrostatic monochromators possess angular aberrations. In order to achieve good monochromaticity and a well-defined transmission curve as function of energy, the angular spread of the input beam must be confined to a few degrees, which causes constraints on the design of the electron emission system. The design of the lens system, on the other hand, depends on the beam parameters delivered by the monochromator, namely the current, the angular spread of the beam, the energy, and the dimensions of the exit slit. The design of an optimum electron source for IPES is therefore a multi-parameter problem, which neither has an exact, nor a unique solution.

The key unit which defines the specifications of the entire source is the electrostatic monochromator. Here, we follow the routes paved in the past with the development of electron energy loss spectroscopy (EELS). For this type of spectroscopy a new electrostatic monochromator of the deflector type was invented by Ibach [15]. The deflector features curved, concave-shaped deflection plates and a total deflection angle of 146°. The 146°-deflector can handle higher currents than the common electrostatic deflectors, e.g., the spherical or the cylindrical deflector. However, so far, the 146°-deflector has been employed merely for EELS where high energy resolution (<5 meV) is required. Monochromatic currents usually range between 0.1 and 1 nA, far less than required for IPES (>100 nA). On the other hand, an energy resolution of 50-100 meV would already represent a substantial improvement for the IPES method. In order to understand whether enough current at moderate resolution may be generated by the presently realized electron sources with the 146°-deflector, it is useful to look at the currents obtained with such equipment as function of the full width at half maximum (FWHM) of the beam energy [16]. The experimental data plotted in Fig. 2 fit to a power law with an exponent 1.9.

According to the theory of space-charge-limited currents between two parallel plates, one might expect an exponent 1.5 for the dependence of the current on the monochromator pass energy and hence on the FWHM of the output current [17]. However, inside the monochromator the charge density distribution is quite different from the parallel plate situation. Firstly, because the electron energy in a monochromator remains nearly constant along the electron path, unlike in the parallel plate geometry. Secondly, in a monochromator the electron density drops dramatically along the path because of the energy dispersion.

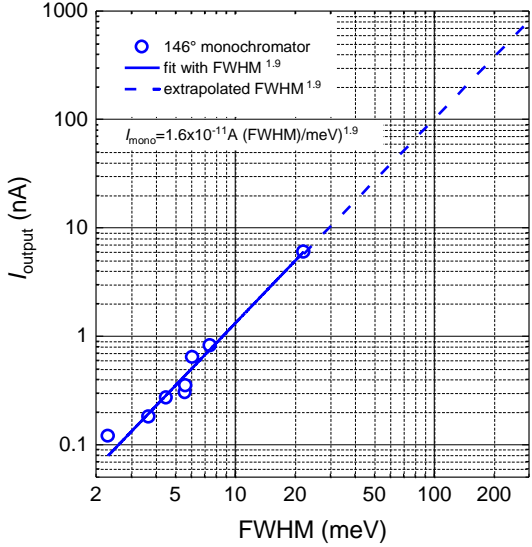

Fig. 2: Experimental data for the current as function of the energy spread of the beam (FWHM) delivered by the EELS electron source described in [16] are depicted as open circles. The dashed blue line shows the extrapolation to the output high currents required for IPES.

The experimental exponent of about two (1.9) (see Fig. 2) is consistent with our simulations of the monochromatic current as function of the FWHM. The blue dashed line in Fig. 2 depicts an extrapolation of the plot into the regime of currents that would be useful for IPES. The extrapolation shows that currents larger than 100 nA can only be obtained at the expense of a FWHM larger than 100 meV.

In this paper we describe an electron source comprising of a more compact electron emission system than used previously [16] in combination with a monochromator which is adapted to high input currents by extending the deflection angle to 162°. Finally, a lens system is constructed which is able to handle the high currents and produces a focused beam of low energy down to 1.6 eV at the target.

The paper is organized as follows. The design of all elements of the electron source is guided by numerical simulations, which consider the electron-electron interaction in a continuum approximation. This approximation is outlined in the next section. Sections 3, 4, and 5 describe the electron emission system up to the entrance slit of the monochromator, the monochromator, and the lens system, respectively. Some key results of the simulations are likewise presented in these sections. Experimental results with respect to monochromator transmission function and monochromatic current are presented in Sec. 6. As expected, we find the energy resolution to degrade at higher currents. Quite unexpectedly, however, we find the energy spread to become smaller again for even higher currents, even smaller than in the low-current case. We attribute this narrowing of the energy spread in the longitudinal direction to binary electron-electron collisions which scatter electrons of different longitudinal velocity out of the beam, whereby these electrons fail to pass the exit slit of the monochromator. In Sec. 7 we present the beam shape as a function of the kinetic energy on the target. The paper concludes with a comparison of the performance of the new system with previous ones.

## 2. Electron-electron interaction in the continuum approximation

The high current in the electron emission system, the monochromator, and to a lesser degree also in the lens system entails that it is indispensable to take the repulsive Coulomb interaction between electrons into account. While there is no analytical solution for the Coulomb many-body interactions except for special beam geometries [12], an approximate numerical solution of the problem is obtained in a scheme in which all other electrons except the one considered are treated as a charge continuum, the so-called *space charge*. To that end, one first calculates the potential distribution inside the system of interest by solving numerically the Laplace equation on a grid of cubes, fine enough to represent the system with sufficient accuracy (linear length 0.1-1 mm). In order to have a solution of the Laplace equation for an arbitrary set of voltages on the electron lens elements, the Laplace equation is solved separately for each independent controllable electrode, which is loaded by the potential of one volt, while all other voltages remain zero. The solution for arbitrary potentials on the lens elements is then obtained by adding the normal solutions for each electrode multiplied with the actual voltage (thereby making use of the linearity of the Laplace equation). To account for the space charge, the system is filled sequentially with about 5000 electrons. The electron trajectories are obtained by stepwise integration. After each time step, the position of the electron is noted, assigned to one of the cubes and added to the previous number of "hits". The space charge is then calculated by scaling the area-integral of the density of hits to the current and the electron energy, so that the resulting numbers represent the space charge density in each cube. The space-charge-induced potential is obtained by solving the Poisson equation with the charge density alone; that is for zero potential on the lens elements. The total potential is then the sum of all individual potentials, scaled to the input current and the lens element potentials. The space-charge-modified trajectories are obtained by using this total potential. In principle, one could then go into further iterations by calculating the space charge again with the potential obtained in the first iteration, and so forth. However, the procedure converges only for moderately low currents, where a single iteration suffices. Stable results are obtained when the transmission function of the device of interest is calculated by stepwise increasing the current using the space charge density obtained in the previous step.

Considering space charge in the continuum approximation sufficed for the quantitative description of resolution and monochromatic currents of previous designs featuring a less compact electron emission system and a 146°-monochromator [16]. Here, however, because of the high currents involved, the continuum approximation fails in two ways. Firstly, the exchange of transverse and longitudinal momentum in the electron emission system gives rise to a broadening of the energy distribution of the electrons downstream, the so-called *Boersch effect* [11-14]. For the electron emission system presented here, e.g., the energy broadening to be attributed to the Boersch effect was measured to range up to 500 meV. Secondly, we found that energy resolution of the new monochromator employed here is significantly better than predicted by simulations in the continuum approximation. The monochromator also tolerates higher currents than expected. A possible reason for this startling effect is discussed in Sec. 6.

## 3. The electron emission system

The present electron emission system was developed in the context of EELS. While details of the design were not published yet, the electron emission system was employed in recent studies on magnon spectroscopy in ultrathin magnetic films [18-20].

The electron emission system consists of a specially shaped Cu block ("repeller") housing the cathode tip, three Cu metal lens elements named A1, A2, and A3, and the CuBe entrance slit of the monochromator. Fig. 3 displays the system with the exception of several pumping holes in repeller and lens elements. The pumping holes serve for a better vacuum near the cathode tip and thereby for a better stability of the emission current. The dimensions are given in Table 1. The cathode is a commercial $LaB_6$ cathode with a 6 μm diameter flat tip and 60° cone angle (Kimball Physics, ES-423E, $LaB_6$ cathode style 60-06). Compared to previous designs [21], the electron emission system is more compact by a factor of two which, according to laminar flow theory of space charge-limited currents, should provide four times higher currents. In practice, the gain in current was even higher, since the electron emission system is better matched to the monochromator slit geometry by shaping the repeller into the form of a groove.

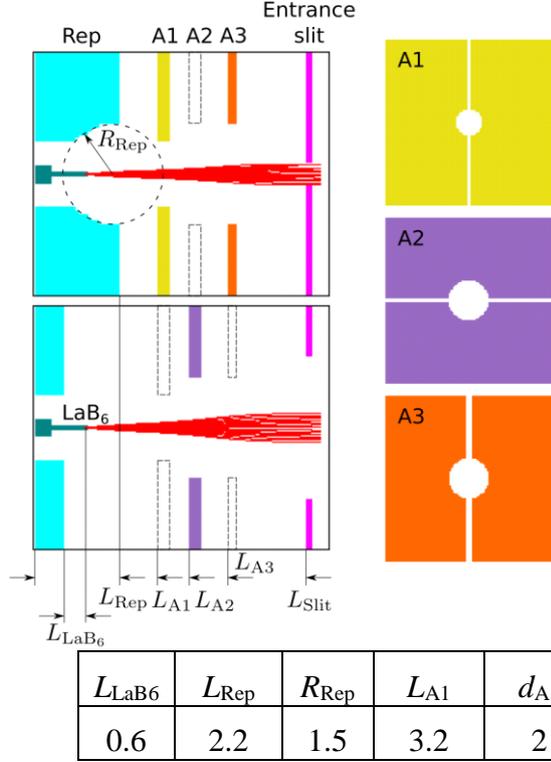

Fig. 3. Sketch of electron emission system together with the definition of distances and radii given in Table 1. The right side shows the profile of the lens element reduced in size to 40% with respect to the left side. The electron trajectories (red lines) are for low emission currents and with the assumption that electrons are emitted into a cone of ±30°. The splitting of the lens elements allows the control of the beam direction into the center of the entrance slit of the monochromator.

| $L_{LaB6}$ | $L_{Rep}$ | $R_{Rep}$ | $L_{A1}$ | $d_{A1}$ | $L_{A2}$ | $d_{A2}$ | $L_{A3}$ | $d_{A3}$ | $L_{Slit}$ |
|---|---|---|---|---|---|---|---|---|---|
| 0.6 | 2.2 | 1.5 | 3.2 | 2 | 4.1 | 3.0 | 5.0 | 3.0 | 7.0 |

Table 1. Dimensions of the electron emission system in mm. See definitions in Fig. 3. $d_{A1}$, $d_{A2}$ and $d_{A3}$ are the diameters of the circular openings of the respective lens elements.

A specific feature of the new electron source, apart from the high current, is that it allows for a control of the angular spread entering the monochromator by merely changing a single potential. This is illustrated with Fig. 4. The data refer to a 0.5×2 mm$^2$ entrance slit of the monochromator held at 24 eV electron energy. The emission current of the cathode is 15 μA. Potentials of cathode tip, repeller, A1, A2, A3, and the slit are 0 V, –7 V, 50 V, 5 V, and 24 V, respectively.

On the left ordinate, the current that passes through the slit ($L_{slit}$ in Fig. 3) is plotted. On the right ordinate the variances $\sigma_\alpha$ and $\sigma_\beta$ of the gaussian distribution of angles relative to the central path at the entrance slit are shown.

Fig. 4 shows that the angular spread is reduced by a factor 3-4 for lower potentials on the A3 lens element at the expense of the delivered current. Especially in the $\beta$-plane the electron trajectories form a nearly parallel beam. According to our calculations, a similar reduction of the angular spread at the expense of the input current is also obtained when voltages either on A1 or on A2 are reduced. The reduction of the angular spread at lower A3 potentials may have a significant effect on the resolution of the monochromator

We remark that the actual variances of angle distributions may differ from the calculations, since the continuum approximation to the space charge problem as employed here does not consider the significant broadening of the energy distribution along the path between cathode tip and entrance slit of the monochromator (see Sec. 2).

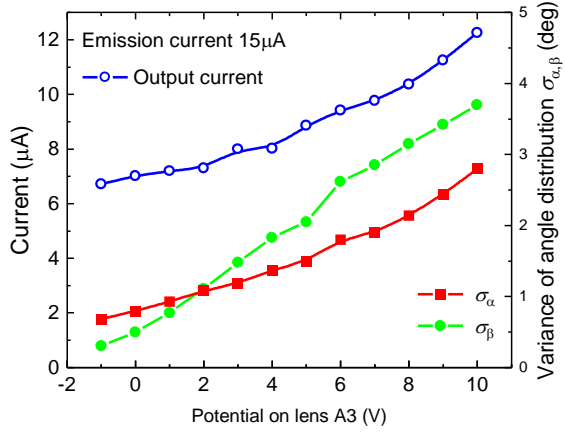

Fig. 4. Left ordinate: calculated output current of the electron emission system into the $0.5 \times 2$ mm$^2$ entrance slit of a monochromator as function of the potential on lens element A3. Voltages at repeller, lenses A1, A2, and the slit are set to $-7$ V, 50 V, 5 V, and 24 V, respectively. The voltages refer to the cathode tip potential. Right ordinate: variances $\sigma_\alpha$ and $\sigma_\beta$ of the gaussian distribution of angles relative to the central path at the entrance slit. The angles $\alpha$ and $\beta$ refer electron trajectories parallel and perpendicular to the dispersion plane of the monochromator, respectively.

## 4. The monochromator

The monochromator is a variant of the monochromator invented for electron energy loss spectroscopy (EELS) [15]. The monochromator features convex-shaped deflecting plates (Fig. 5a, b), which, in combination with a negative potential on the top and bottom cover plates, lead to curved equipotential lines (Fig. 5b). If the total deflection angle is 146° and a particular potential on the top and bottom cover plates is applied, the curved equipotential lines provide stigmatic focusing at the exit as in a spherical deflector. However, contrary to the spherical deflector, the radial electric field decays in intensity on both sides of the central path, which reduces the angular aberration in the dispersion plane. Furthermore, the stigmatic focusing results from the shape of the equipotential lines, not from symmetry as in the spherical deflector. The focusing properties are therefore less susceptible to space-charge-induced distortions and the Ibach monochromator can therefore handle much larger currents than other electrostatic deflectors, such as the spherical deflector [22].

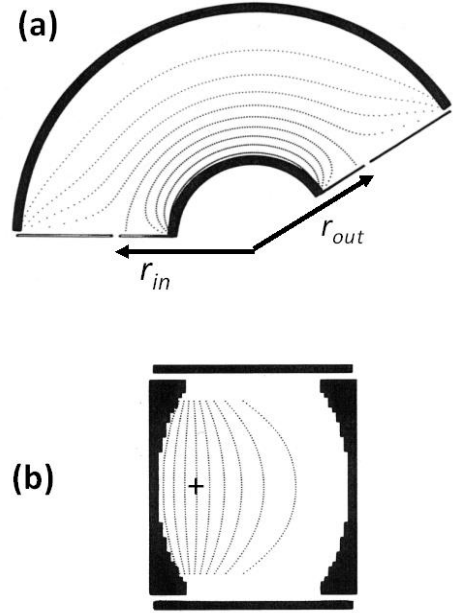

Fig. 5. The 146°-electrostatic deflector analyzer. (a) Top view: the solid black lines are slits and deflector plates in the center plane, the dotted lines are equipotential lines. The radial positions of entrance and exit slits are marked as $r_{in}$ and $r_{out}$; (b) Side view: the deflecting plates have a convex shape (radius 100 mm), which, in combination with a negative potential on top and bottom cover plates, leads to curved equipotential lines. The cross marks the approximate position of the central path (after Ref. [16]).

| Slits | Radial position | Width | Height |
|---|---|---|---|
| Entrance | 37 mm | 0.5 mm | 2 mm |
| Exit | 39 mm | 0.5 mm | 3 mm |

Table 2: Dimensions and radial positions of the entrance and exit slits ($r_{in}$ and $r_{out}$). The slightly larger radial position at the exit ensures that the distribution of exit angles $\alpha$ centers at $\alpha \cong 0°$ for high current loads.

Even higher currents can be handled when the deflection angle is extended, as explained in the following. The reason is qualitatively explained for a specifically designed 162°-deflector with dimensions as displayed in Table 2.

For a deflection angle of 162°, one may still have a focus in the dispersion plane (radial focus) by adjusting the potential of the cover plates (Fig. 6a). Perpendicular to the dispersion plane the focal point occurs then at about 125° if the input current is small (Fig. 6b).

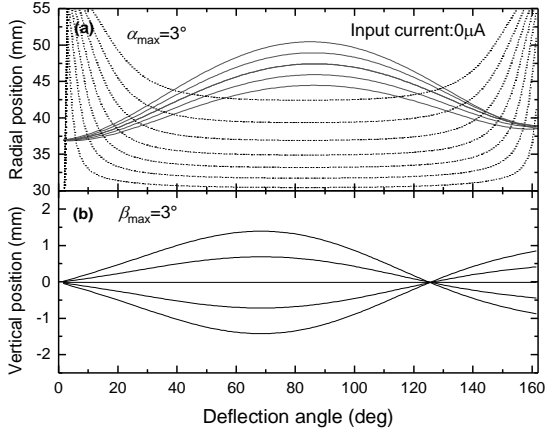

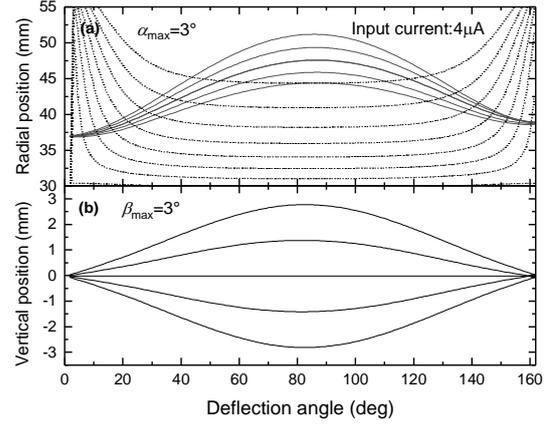

Fig. 6: Solid lines are trajectories of electrons starting from the center of the entrance slit with angles $\alpha$ and $\beta = 0°$, $\pm 1.5°$, and $\pm 3°$ for the case of zero input current. (a) Focus in the dispersion plane is achieved by applying a proper potential to the top and bottom cover plates. (b) Perpendicular to the dispersion plane the focal point occurs at a deflection angle of about 125°. Dashed lines are equipotential lines.

For larger input currents with the potential of the cover plates adjusted to obtain the radial focus still at the exit slit, the vertical focus, by virtue of the space-charge-induced repulsion, shifts to larger deflection angles. For an input current of 4 µA, one obtains a stigmatic focus at the exit slit (Fig. 7), as for the 146°-deflector in the low current limit. The deflector with an extended deflection angle is therefore optimized for a particular, nonzero input current.

Fig. 7: Solid lines are trajectories of electrons starting from the center of the entrance slit with angles $\alpha$ and $\beta = 0°$, $\pm 1.5°$, and $\pm 3°$, now for 4 µA input current. (a) Focus in the dispersion plane is achieved by applying a proper potential to the top and bottom cover plates. (b) Perpendicular to the dispersion plane the focal point occurs now also at a deflection angle of 162°. Dashed lines are equipotential lines.

Fig. 8 shows the FWHM of the simulated transmission function of the 162° deflector as function of the input current for two different variances of the input angle distribution. The entrance slit is now loaded with a homogeneous distribution of electron trajectories.

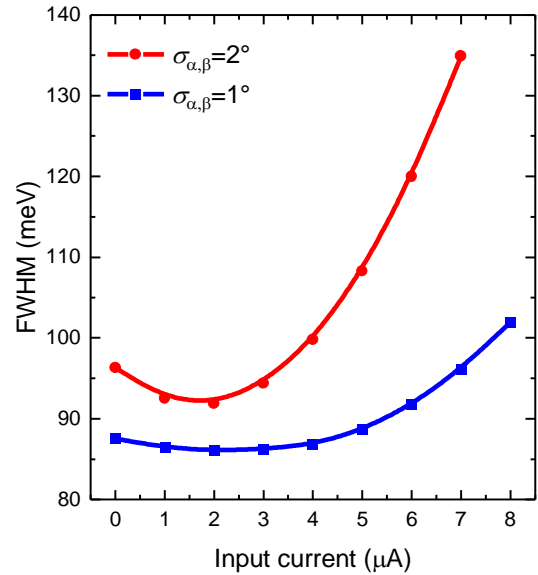

Fig 8: FWHM of the transmission functions obtained from simulations for two different variances of the gaussian distribution of angles $\alpha$, $\beta$. Entrance and exit slit dimensions are as in Table 2.

As to be expected from the theoretical considerations above, the minimum FWHM is not obtained at zero input current but at a current of about 2-3 µA.

## 5. The lens system

The final task is now to bring the monochromatic current to the surface of a target. Moreover, the lens system must be designed to fit into the existing chamber and mirror design of H. Yoshida [23]. The center hole in the mirror limits the diameter of the lens elements. Furthermore, the lens system must not block the light.

A number of lens systems obeying these geometrical constraints were studied. The lens system that worked best is sketched in Fig. 9. The lens element B1 next to the exit slit of the monochromator is on the same potential as the exit slit. The region between the monochromator exit slit and the aperture in the center of lens element B1 is therefore field-free. Hence, the electron trajectories are straight lines. The aperture in the center of lens element B1 serves to cut off electrons on trajectories with large angles $\alpha$ with respect to the center path. The typical $\alpha^2$-aberration errors leading to high energy tails in the transmission function of the monochromator are thereby avoided.

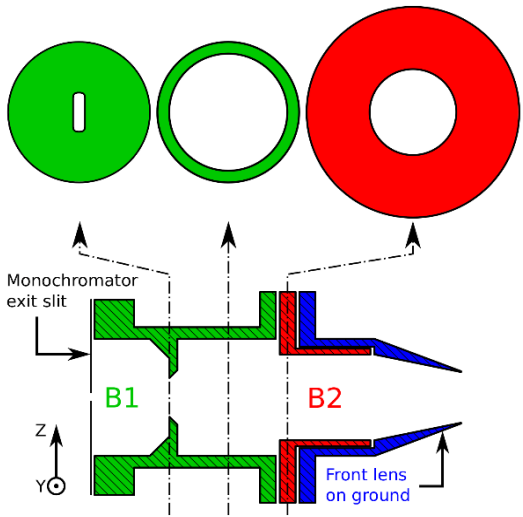

Fig. 9. Sketch of the lens system.

The front lens element facing the target is on the same potential as the target in order to ensure a field-free region around the sample (save for the space charge potential).

Different energies at the target are obtained by varying the potential of all lens elements simultaneously, while the target remains on ground potential. An electron energy of 1 eV at the target, e.g., is obtained if the target is on –23 V with reference to the exit slit of the monochromator when the deflection voltage is 40 V (24 eV pass energy). The focus at the target is accomplished with the suitable potential on the second lens element (B2). For a deflection voltage of 40 eV in the monochromator, good focusing is achieved when the potential is 2.1 V negative with respect to the potential of the exit slit of the monochromator.

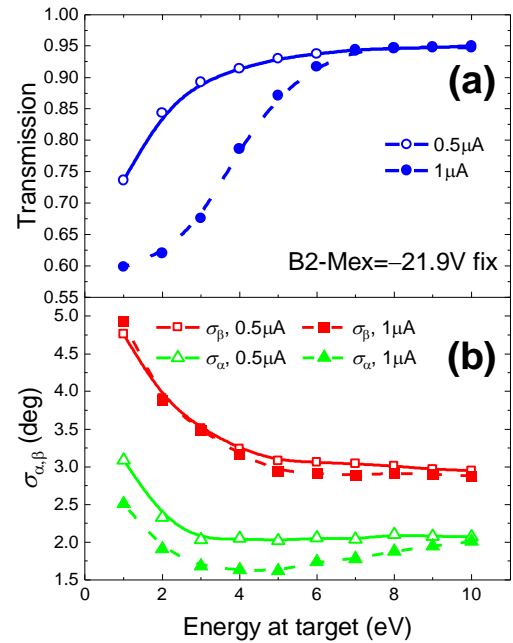

Fig. 10. (a) Ratio of the calculated current arriving at a 3×8 mm² target to the current leaving the monochromator ("transmission") as function of the beam energy at the target. The pass energy of the monochromator is 24 eV. Voltage at lens element B2, minus the voltage at the exit slit of the monochromator, denoted as B2-Mex, is −21.9 V. Solid line is for 0.5 µA current, the dashed line for 1 µA. (b) Variance of the gaussian angle distribution in the dispersion plane $\sigma_\alpha$ and perpendicular to it $\sigma_\beta$ vs. beam energy at the target. B2−Mex is as in (a). Solid lines are for 0.5 µA current, the dashed lines for 1 µA.

Fig. 10 shows the calculated properties of the lens system drawn in Fig. 9. In the simulations the target is assumed to have the dimensions 3×8 mm$^2$. For an 8×8 mm$^2$ rotated sample all electrons hit the surface up to a rotation angle of 67°.

The variances of the angle distributions of the beam leaving the monochromator are assumed to be $\sigma_{\alpha,\beta} = 2°$, in agreement with the simulations for the electron emission system and the monochromator.

The ratio of the current arriving at the target to the current leaving the monochromator ("transmission") for two different currents as function of the kinetic energy at the target is plotted in Fig. 10a by solid and dashed lines for 0.5 μA and 1.0 μA current, respectively. A maximum transmission of about 95% is achieved. The transmission becomes lower as the energy at the target is lower, mostly because the space charge potential drives the electron trajectories apart. Higher currents lead to a larger degradation of the transmission at low kinetic energies at the target (dashed and solid lines in Fig. 10a). For a fixed electron energy, a somewhat higher transmission can be accomplished by a small variation of the potential on lens element B2 (see Sec. 7.1).

Fig. 10b shows the variances of the angle distribution $\sigma_{\alpha,\beta}$ of electrons arriving at the target. These numbers are important for the calculation of the distribution of electron momenta parallel to the surface. The angle distributions with and without current load differ only marginally. The variance $\sigma_\beta$ for the angle $\beta$ is larger than the variance $\sigma_\alpha$, since electrons leaving the slit at a height-position near the upper or lower edge of the slit embark on trajectories which are significantly off the optical axis. Focusing these electrons onto the target leads to larger angles of incidence at the target.

## 6. Experimental results: transmission function

To characterize the electron source, we measure (i) the total current, (ii) the beam shape and size, (iii) the transmission function (FWHM and shape), all as function of the kinetic energy at the target, and (iv) the input current. For (i) and (ii), a Faraday cup is mounted on a manipulator and placed on the optical axis of the electron source. The cup has two openings of 1 mm$^2$ and 2×8 mm$^2$ at different positions. The first one is used to determine the beam profile with 1-mm resolution. The second one is used to measure the total current. For (iii), the Faraday cup can be completely removed out of the target area in order to measure the FWHM and shape of the transmission function vs. energy, using a single 146°-analyzer [16] placed in front of the electron source. For (iv), as input current we take the current measured on the outer lens of the monochromator when the deflection voltage is reversed (e.g., -40 V instead of 40 V). By doing so we make sure that all electrons that pass the entrance slit are collected.

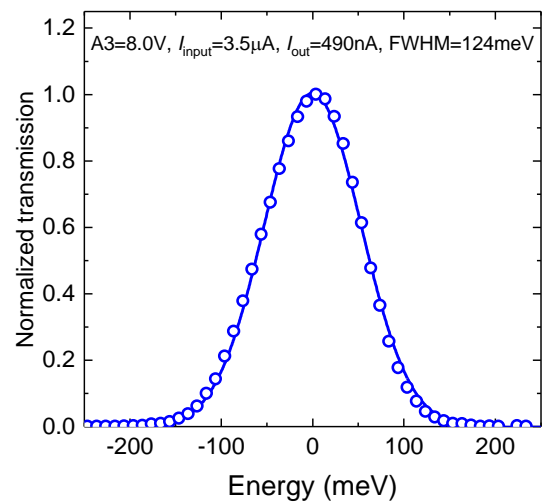

Fig. 11: Measured normalized transmission curve of the 162°-deflector when fed with 3.5 μA input current. The deflection voltage is 40 V. The filament is heated with a current of 1.35 A. A3 is set to 8.0 V with reference to the cathode tip. Output current is 490 nA, the FWHM is 124 meV.

Fig. 8 predicts that the minimum FWHM should be obtained for an input current between 2 μA and 4 μA. This is consistent with Fig. 7 showing that the small-angle focus of a beam emerging in the center of the entrance slit is at the exit slit when the input current is 4 μA. Fig. 11 shows the measured transmission function for 3.5 µA input current. The deflection voltage of the monochromator is 40 V. The filament heating current is 1.35 A. The shape of the transmission function in Fig. 11 is well represented by a gaussian (solid line).

For input currents larger than about 4µA the simulations predict a rapid increase in the FWHM (Fig. 8) and furthermore a gross distortion of the transmission function, e.g. by side wings. Contrary to these predictions, the experimental transmission functions remain gaussians and have an even smaller FWHM at higher input currents (Fig. 12). Again, the deflection voltage is 40 V. Repeller, A1, and A2 are set to –9 V, 61 V, and 18 V, respectively, and A3 is set to + 3.8 V and +9.8 V, respectively. Reference for the potentials is the cathode tip. The maximum output current of 970 nA is obtained for +9.8 V on A3 and a filament current of 1.55 A. The transmission function (red squares in Fig. 12) has a FWHM of 97 meV. The input current into the monochromator is 9.0 µA, three times higher than the optimum current suggested by the simulations (Fig. 8). Even smaller FWHM may be obtained by reducing the potential on A3, while all other potentials remain as before. As can be seen from Fig. 4, the lower potential on A3 narrows distribution of angles in the input beam of the monochromator. For an A3 potential of +3.8 V, the FWHM is 78 meV; input and output currents are 11.9 µA and 664 nA, respectively (blue circles in Fig. 12). Both transmission functions in Fig. 12 are well described by gaussians (solid lines) with low intensities in the tail, whereas numerical simulations in the continuum approximation yield broad energy distributions with extreme tails on the low energy side.

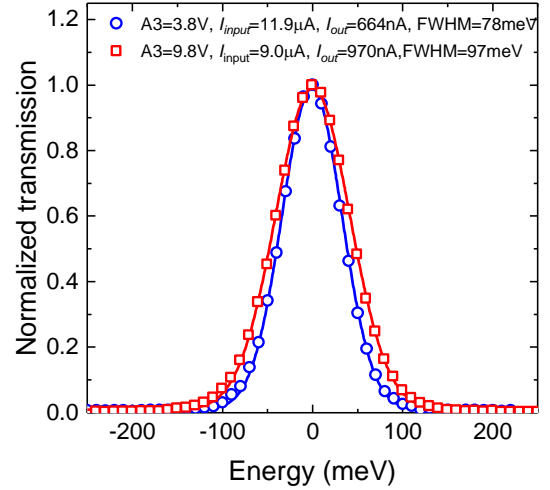

Fig. 12: Measured normalized transmission functions of the 162°-deflector when fed by input currents of 9 µA and 11.9 µA. The filament current is 1.55 A. The deflection voltage is 40 V as for Fig. 11.

Hence, the experimental results fall into two different regimes: the low input current regime, in which simulations and experimental result agree. Fig. 11 represents an example of this. However, even lower FWHM can be obtained for a factor of 3-4 higher input currents (Fig. 12).

As a qualitative interpretation for this remarkable effect we propose a mechanism of velocity selection, which considers the particle nature of electrons: Inside the monochromator, electrons essentially continue to travel with the longitudinal speed by which they have entered the entrance slit. Electrons with a particular longitudinal speed $v_{pass}$ embark on those trajectories which run through the exit slit. Electrons with a speed rather different from $v_{pass}$ leave the central region of the beam due to the velocity dispersion of the monochromator. Electrons whose speed deviates little from $v_{pass}$ may experience repulsive coulomb interactions with other individual electrons having about the same, but moderately different speed. As a result of such mostly noncentral repulsions, electrons may be scattered out of the central region of the beam and would therefore not pass the exit slit. Mainly electrons with about the same speed survive in the central region of the beam. The result is a narrowing of the velocity distribution of those electrons which do pass the exit slit. Hence the result for electrons in the focusing deflecting field is the opposite of the normal Boersch effect, which leads to a broadening of the energy distribution. We therefore denote this effect as the "*inverse Boersch effect*". According to the qualitative description above, the narrowing of the energy distribution should be the larger the higher the electron density and the smaller the angle distribution is. This is consistent with the observation that the smallest FWHM are observed for input high currents (Fig. 12 vs. Fig. 11) and that the FWHM is lowest for low A3 potentials when the angular spread of the beam is small (Fig. 4 and 11).

Numerical calculations which consider electron-electron scattering in dense electron beams have been performed by Read and Bowring [24], however, only for a focused circular beam with no energy selective deflection field and apertures. We note, however, that the effect of velocity selection via removing particles from the beam which have engaged in noncentral binary collisions is well-known for dense He-atom beams emerging from a high-pressure cell via a nozzle [25-28]. In case of He-beams, differential pumping of chambers separated by sheets with small orifices take care that only He-atoms with nearly the same speed in the forward direction survive in the beam. The high monochromaticity of such He-beams have been exploited, e.g., to study inelastic scattering from phonons [29].

## 7. Experimental results: Kinetic energy dependence

For IPES, the kinetic energy of the electron beam at the target is an essential parameter. Therefore, we now explain how to optimally adjust the potentials in the lens system in order to maximize the target current while sweeping the kinetic energy. We then show how the beam shape evolves as a function of kinetic energy.

### 7.1. Lens potentials:

When the resolution of the monochromator is altered, or while sweeping the kinetic energy of the electrons at the target, potentials on the lens elements (Fig. 9) need to be changed. To determine the optimum potentials experimentally, we position a Faraday cup with a 2×8 mm² rectangular opening at the target position. The long axis of the rectangular opening is parallel to the long axis of the monochromator exit slit. For a given deflection voltage on the monochromator $\Delta V$, the current passing through the Faraday cup opening is optimized by varying the voltage on lens element B2 (Fig. 9) according to experimental relation:

$$U_{B2} = c_1 + 0.078 E_{kin}/e, \quad (1)$$

in which $c_1$ depends on the pass energy of the monochromator. Hereby the lens element B1 is on the monochromator exit slit potential and the front lens element on the target potential. For a pass energy of 24 eV ($\Delta V$ = 40 V), e.g., $c_1$ is −22.5 V. In Fig. 10 we had kept $U_{B2}$ fixed at −21.9 V whereas according to Eq. 1 $U_{B2}$ should vary between −22.4 V and −21.1 V for a variation of the target kinetic energy between 1 eV and 10 eV.

When changing the deflection voltage $\Delta V$ of the monochromator, and thus the energy resolution, one needs to adjust the mean voltage of the inner and outer deflection plate $U_M$ as well as the mean voltage on the cover plates $U_{Cov}$ of the monochromator. When $\Delta V$ is varied by the amount $d\Delta V$ then the mean voltage of the monochromator should vary by

$$dU_M = 0.124 \text{ V} \times d\Delta V \quad (2)$$

and the mean voltage of the cover plates by

$$dU_{Cov} = 0.607 \text{ V} \times d\Delta V, \quad (3)$$

as found experimentally.

Neglecting aberrations, the energy resolution of the electron source should be proportional to the deflection voltage $\Delta V$. That is, however, not the case. Because of phase space conservation the variance of the gaussian distribution of angles $\sigma_\alpha$ scales with the pass energy $E_{pass}$ as $E_{pass}^{-1/2}$. The α-angle aberration in the monochromator is quadratic in α. Experimentally, the FWMH of the energy resolution is therefore approximately described by a constant plus a linear term in $\Delta V$:

FWHM = 9.6 meV+2.24 10⁻³ $e$ $\Delta V$.    (4)

### 7.2. Beam profiles

In Fig. 13 we show the electron beam profiles measured with a circular opening of 1 mm for kinetic energies at the target between 2.3 and 21.3 eV. The Y- and Z-directions are perpendicular and parallel to the long direction of the exit slit of the monochromator, respectively (see Fig. 9).

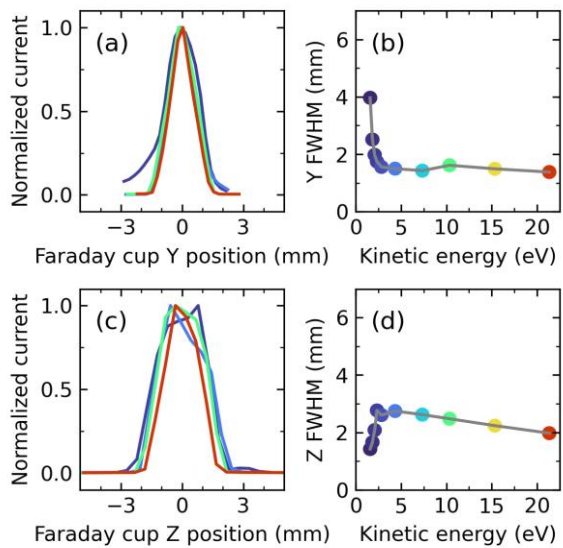

Fig. 13: (a) Beam profiles perpendicular to the long side of the exit slit (Y axis in Fig. 9) of the monochromator for 2.3 eV, 4.3 eV, 10.3 eV, and 21.3 eV vs. kinetic energy at the target. (b) FWHM of the profiles in (a) vs. kinetic energy at the target (from 1.6 eV to 21.3 eV). (c) Beam profiles parallel to the long sides of the exit slit (Z axis in Fig. 9) of the monochromator for 2.3 eV, 4.3 eV, 10.3 eV, and 21.3 eV kinetic energy at the target. (d) FWHM of the profiles in (c) vs. kinetic energy at the target.

The deflection voltage of the monochromator is 20 V, the output current is 306 nA, the FWHM of the transmission function is 54 meV and the filament current is 1.55 A. The distortion of the shape of the beam at low kinetic energies shown in Fig. 13b and 13d is due to space charge build-up in the lens at low energies. For smaller currents the shape distortions are reduced which results in a higher transmission (Fig. 10).

### 8. Summary of results

Fig. 14 shows the measured output current as function of the FWHM of the electron beam. The blue dashed line represents previous results extrapolated to higher currents (Fig. 2), the symbols mark current and resolution of the present system for deflection voltages of 10 V, 20 V, 40 V, 50 V, and 60 V. The two blue dots are both for 40 V deflection voltage, however for different values of the A3-potential (Fig. 12). Scaled to the same FWHM, currents of the new system are an order of magnitude higher than for the previous EELS electron sources.

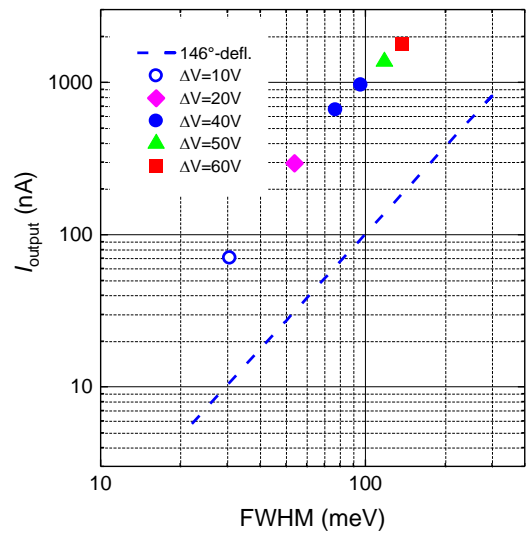

Fig. 14: Output current vs. FWHM for deflection voltages $\Delta V$=10 V, 20 V, 40 V, 50 V, and 60 V are shown as a blue open circle, a magenta diamond, two blue dots, a green triangle, and a red square, respectively. The filament current is 1.55 A. The two blue dots are for the two different potentials on A3 (Fig. 12). The blue dashed line is the same line as in Fig. 2 and represents data for the 146°-deflector extrapolated to higher currents.

The design of the new 162°-monochromator was guided by simulations. However, the experimental results are significantly better than predicted by the simulations, which neglect binary electron-electron collisions. Because of that, we presently cannot tell whether or not, and possibly how, further improvements may be conceivable.

**Author's contributions:**

H.Y. and F.C.B. conceived the original idea. H.I., F.S.T., H.Y., and F.C.B. planned and implemented the research. H.I. designed the electron optics based on programs which include electron/electron interaction in the continuum approximation. H.I. implemented the extension of the deflecting angle of the monochromator in order to accommodate higher currents. H.I. H.S., M.K., and F.C.B. assembled the electron source. F.C.B. performed the experimental measurements and H.S. helped the measurements. F.C.B. developed the control software. H.I. and F.C.B. produced the figures. H.I. wrote the manuscript with the help of all authors.

**Acknowledgments**

H.I., F.S.T., and F.C.B. acknowledge the financial support of the DAAD, project 57524906. F.S.T. and F.C.B. acknowledge financial support by the Deutsche Forschungsgemeinschaft (DFG), Project-ID 223848855-SFB 1083. JST SPRING grant (JPMJSP2109). H.S, M.K. and H.Y. thank JSPS researcher exchange program (JPJSBP120203504). H.Y. acknowledges the financial support of JSPS KAKENHI (21H05472 and 21H01902).

**Data Availability Statement**

The data that support the findings of this study are openly available at the Jülich DATA public repository:

https://doi.org/10.26165/JUELICH-DATA/9TAJ4D

**Conflict of Interest**

The authors have no conflict of interests to disclose.